\documentclass[10pt]{iopart}
\usepackage{iopams}

\expandafter\let\csname equation*\endcsname\relax

\expandafter\let\csname endequation*\endcsname\relax

\usepackage{amsmath}

\usepackage{graphicx}
\usepackage{bm}
\usepackage{etoolbox}
\usepackage{physics}
\usepackage{stackengine}
\usepackage{mathtools}
\usepackage{lipsum}
\usepackage{breqn}
\usepackage{hyperref}
\usepackage{nicefrac}
\usepackage{dblfloatfix} 
\usepackage{afterpage}
\hypersetup{
    colorlinks=true,
    linkcolor=blue,     
    urlcolor=cyan,
    citecolor=blue,
}

\AtBeginEnvironment{align}{\setcounter{subeqn}{0}}
\newcounter{subeqn} \renewcommand{\thesubeqn}{\theequation\alph{subeqn}}%
\newcommand{\subeqn}{%
  \refstepcounter{subeqn}
  \tag{\thesubeqn}
}

\begin{document}

\title[Hard nonlinear evolution in the presence of particle orbits]{Impact of energetic particle orbits on long range frequency chirping of BGK modes}

\author{H. Hezaveh$^{1,2}$, Z. S. Qu$^{2}$, B. Layden$^{2}$ and M. J. Hole$^{2}$}

\address{$^1$ Faculty of Sciences and New Technologies, University of Isfahan, Isfahan, Iran}
\address{$^2$ Research School of Physics and Engineering, The Australian National University, Canberra ACT 2601, Australia}
\ead{hooman.hezaveh@anu.edu.au}
\date{today}

\begin{abstract}

Long range frequency chirping of Bernstein–-Greene–-Kruskal modes, whose existence is determined by the fast particles, is investigated in cases where these particles do not move freely and their motion is bounded to restricted orbits. An equilibrium oscillating potential, which creates different orbit topologies of energetic particles, is included into the bump-on-tail instability problem of a plasma wave. With respect to fast particles dynamics, the extended model captures the range of particles motion (trapped/passing) with energy and thus represents a more realistic 1D picture of the long range sweeping events observed for weakly damped modes, e.g. global Alfven eigenmodes, in tokamaks. The Poisson equation is solved numerically along with bounce averaging the Vlasov equation in the adiabatic regime. We demonstrate that the shape and the saturation amplitude of the nonlinear mode structure depends not only on the amount of deviation from the initial eigenfrequency but also on the initial energy of the resonant electrons in the equilibrium potential. Similarly, the results reveal that the resonant electrons following different equilibrium orbits in the electrostatic potential lead to different rates of frequency evolution. As compared to the previous model [Breizman B.N. 2010 Nucl. Fusion 50 084014], it is shown that the frequency sweeps with lower rates. The additional physics included in the model enables a more complete 1D description of the range of phenomena observed in experiments.

\end{abstract}

%
%
%
%
\ioptwocol

\section{Introduction}
\label{sec:intro}

Fast particles are abundantly present in burning plasmas. They exist either through external heating or eventually by fusion-born alpha particles. Energetic particle driven modes (EPMs) \cite{chen} can occur as a result of fast particles interaction with weakly damped plasma modes, e.g. Alfven eigenmodes (AEs) \cite{Heidbrink}. The resulted excited modes can cause the undesirable ejection of energetic particles from the hot core towards the walls of a toroidal machine \cite{Iterphysicsexperts,Fasoli2007,Heidbrinkreview}. This loss deteriorates plasma heating and degrades the confinement in a power plant. Accordingly, understanding the behavior of these modes is momentous in burning plasmas of future fusion reactors.

Experimental results, in the case of neutral beam injection, demonstrate that EPMs, as a result of AEs excitation, exhibit a “hard” nonlinear regime \cite{HeidbrinkDIIID,Wong,sdpinches,Fredrickson2006a,Shinohara} with rapid frequency sweeping. Small deviations from the initial eigenfrequency for the case of a near-threshold instability $\abs {\gamma_{l}-\gamma_{d}} \ll \gamma_{d} \le \gamma_{l}$, where $\gamma_{l}$ is the kinetic drive and $\gamma_{d}$ is the damping rate due to dissipation in the background plasma, were first studied using a 1D bump-on-tail (BOT) model by Berk-Breizman (BB) and co-workers \cite{Berk1997}. This model shows the nonlinear process of holes and clumps formation in the fast particle distribution function. A pair of Bernstein-–Greene–-Kruskal (BGK) \cite{BGK} nonlinear modes chirping up and down in frequency is supported by these nonlinear phase-space structures and the frequency shifts are associated with the motion of these coherent structures due to energy dissipation in the bulk plasma. The much longer evolution time scale of these nonlinear structures in comparison with their development time scale in the explosive formation stage is one of the key results in \cite{Berk1997} to be taken into consideration. It should be mentioned that holes and clumps form not only in case of a weakly unstable mode but also with any amount of background dissipation \cite{Lilley2014}. The Berk-Breizman scenario has been proved to be successful in explaining the frequency chirping events observed in experiments with AEs \cite{Fasoli1998,Heeter}. Moreover, the effect of different types of relaxation processes on the nonlinear evolution has been investigated in \cite{Lilley2009} and \cite{Lilley2010}, with the BOT code introduced in the latter. All the mentioned models are based on the assumption that the range of frequency chirping is short and the mode structure is fixed.  

However, experimental evidence exists for mode activities in which the frequency shifts are as large as the initial eigenfrequency itself \cite{Gryaznevich,Maslovsky,Fredrickson2006b}. As the mode amplitude saturates due to flattening of the distribution function of the energetic particles, the physical picture of each evolving phase-space structure is a BGK mode whose frequency changes in time and its structure is notably affected by the frequency shift. Recently, a nonperturbative model based on the adiabatic description of the fast particles contribution has been developed by Breizman \cite{Boris2010} using a 1D BOT instability to interpret the long range chirping for an isolated nonlinear resonance. This approach is premised on the assumption that the width of the separatrix supported by the BGK mode is small compared with the characteristic width of the unperturbed distribution function. The Breizman model remains valid as long as the separatrix of the energetic particles inside the clump shrinks for a downward shift in the frequency. As an extension, the adiabatic description of treating an expanding separatrix which traps the ambient particles is presented in \cite{Nyqvist2013} by Nyqvist and Breizman.

\begin{figure}[t]
  \centering
   \includegraphics[scale=0.55]{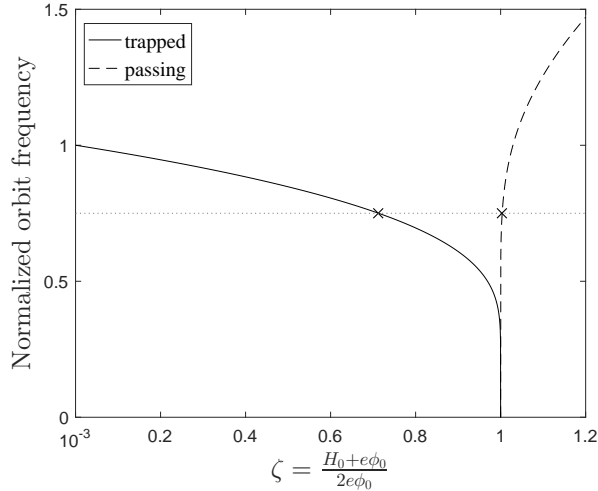}
  \caption{Normalized equilibrium frequency of the fast particles in the electrostaic potential vs. energy parameter. The dotted line shows a sample eigenfrequency simultanously in resonance with particles of two different orbit types}
  \label{fig1}
\end{figure}

In magnetized plasmas, e.g. magnetic confinement devices, the particles gyrate about the magnetic field lines and follow certain trajectories depending on their energy and the magnetic field inhomogeneity. Therefore, the impact of particle orbits on the long range frequency sweeping events, should also be investigated in order to better understand and control these instability-driven phenomena. An electrostatic model where the energetic particles are not moving freely and their motion is bounded to certain orbits, enables such an investigation through a 1D picture. This physical model is the subject of this paper. We add a fixed equilibrium oscillatory electrostatic potential to the BOT problem presented in \cite{Boris2010}, thus creating an energy-dependence of the particle oscillation frequency in this equilibrium potential. In this new model, the unperturbed motion of the fast particles in the equilibrium electrostatic potential is governed by the following Hamiltonian 
\begin{equation}
H_{0}=\frac{p_{z}^2}{2m}-e\phi_0 \cos(k_{\text{eq}}z),
\label{eq:equilibriumhamiltonian}
\end{equation}
where $p_z$ is the momentum of the fast particles, $m$ the particle mass and $\phi_0$ and $k_{\text{eq}}$ are the amplitude and wave-number of the equilibrium potential, respectively. The energetic particles interacting with the perturbed field, are considered as trapped or passing in this equilibrium potential, depending on their energy with respect to the electrostatic potential energy. \Fref{fig1}, whose construction is detailed at the end of Subsection \ref{sec:dynamics}, demonstrates the behavior of the equilibrium oscillation frequency of the fast electrons versus their energy. For each frequency of trapped particles motion in the equilibrium potential, there exist a group of passing particles having the same frequency of the motion. Hence, the mode can be simultaneously in resonance with both the trapped and passing electrons in this equilibrium potential. This trapped and passing locus model resembles the trapped particles following the banana orbits and the passing particles in the magnetic field lines of a tokamak (Cf. Section \ref{sec:conclusion}). In addition to enabling the impact of particle orbits on the long range chirping of BGK modes, the contribution from different resonances can also be investigated through the energy dependence. 
 
The nonlinear wave equation is expanded using Fourier decomposition which allows us to find an explicit expression for the Hamiltonian of the fast particles motion in terms of the action-angle variables of the unperturbed motion. This expansion, together with treating the kinetic equation adiabatically, allows us to implement a numerical treatment to investigate the impact of particle orbits on the structure and the sweeping rate of the nonlinear wave. 

In Section \ref{sec:model}, the basic system of equations adopted for the analysis and the dynamic equations of the unperturbed motion is presented, followed by the derivation of the linear growth rate, the equation for the BGK mode structure and the chirping rate. The numerical scheme used for solving the equations is assigned to Section \ref{sec:numeric}. Section \ref{sec:result} presents the results in the regions where the adiabatic invariant of the trapped particles in the BGK mode decreases (the separatrix shrinks) during chirping and the effect of the electrons equilibrium orbit on the nonlinear evolution of the mode. Finally, Section \ref{sec:conclusion} contains concluding remarks.

\section{The model}
\label{sec:model}

In this extended 1D BOT model, we study a purely electrostatic mode in a plasma consisting of static background ions, cold electrons responding linearly to the mode and fast electrons which are trapped and co/counter-passing in a fixed equilibrium oscillatory potential and are in resonance with the electrostatic mode. Cold electrons and ions do not interact with the equilibrium potential. The distribution function of each group of the fast electrons, which is treated through the Vlasov equation, is assumed to be a linearly increasing function of the fast particles energy providing the instability drive. The damping mechanism is the friction force acting on the cold electrons, a necessary condition in this model for the formation of a nonperturbative BGK mode. The system of equations considered to investigate both the linear evolution of the mode and the structure of the BGK mode during frequency chirping consists of Poisson, Vlasov, equation of motion and continuity equations, given by
\begin{align}
&\frac{\epsilon_0}{e}\frac{\partial^2 U}{\partial z^2}=-e\left[\sum_{\alpha} \int \tilde{f}_{\alpha} dv +\delta n \right], \label{eq:systema} \refstepcounter{equation}\subeqn \\
&\frac{\partial f_{\alpha}}{\partial t}+\{f_{\alpha},H\}=0,\label{eq:systemb}  \subeqn \\
&\frac{\partial V_c}{\partial t}=-\frac{1}{m_e}\frac{\partial U}{\partial z}- \nu V_c, \label{eq:systemc} \subeqn \\
&\frac{\partial \delta n}{\partial t}=-n_0 \frac{\partial V_c}{\partial z},  \label{eq:systemd} \subeqn
\end{align}
with $\alpha$ a label that denotes the orbit type of the fast electrons motion in the electrostatic potential: $ \left (\alpha=t \right )$ and $ \left (\alpha=p \right )$ for the trapped and passing electrons in this potential, respectively. The total distribution function is $f_{\alpha}=F_{\text{eq},\alpha}+\tilde f_{\alpha}$, with $F_{\text{eq},\alpha}$ and $\tilde f_{\alpha}$ being the initial and the perturbed parts, respectively. The energy of the electrostatic mode is given by $U$, $\epsilon_{0}$ is the permittivity of free space, $m_e$ the electron mass, $e$ the electron charge, $\nu$ the collision frequency of the cold electrons, $V_c$ the flow velocity of the cold electrons and $n_0$ and $\delta n$ the unperturbed and perturbed density of the cold electrons, respectively.

\subsection{Fast particles orbits and dynamics}
\label{sec:dynamics}

For the completely integrable system consisting of trapped and co/counter-passing particles whose motion is goverened by the Hamiltonian presented in equation \eqref{eq:equilibriumhamiltonian}, it is possible to transform canonically from the variables $\left(z,p_z\right)$ to action--angle variables $\left(\theta,J_\alpha\right)$, written as
\begin{align}
J_{\alpha=t}= \frac{2}{\pi}\int_{0}^{z_{\text{max}}}p_z dz  = \frac{8\sqrt{m_e \phi_0 e}}{k_{\text{eq}}\pi} \nonumber \\
\ \ \ \ \ \ \ \ \ \times \left [ (\zeta-1 ) K (\zeta ) + E  (\zeta \right ) ],   \refstepcounter{equation}\subeqn \\
J_{\alpha=p}=\frac{1}{2\pi} \int_{0}^{\lambda} p_z dz = \frac{4 \sqrt{e \phi_{0}}  }{k_{\text{eq}} \pi} E \left ( \zeta \right ), \subeqn
\end{align}
where $J_{\alpha}$ is the action for the unperturbed motion of the fast particles, $z_{\text{max}}$ is determined by $p_z=0$ using equation \eqref{eq:equilibriumhamiltonian}, $\lambda$ is the wavelength of the equilibrium potential, $\zeta$ is the energy parameter (representing the electron orbits) equal to $\frac{H_0 \left (J_{\alpha}\right ) + e\phi_0}{2e\phi_0}$ with $H_0 \left (J_{\alpha}\right )$ being the unperturbed Hamiltonian written in terms of the action--angle variables and $K(\zeta)$ and $E(\zeta)$ are the complete elliptic integral of the first and second kind, respectively.
Using the canonical equations of motion, the frequency of the motion reads
\begin{align}
&\Omega_{\alpha=t}=\frac{\partial H_{0,\alpha=t}}{\partial J_{\alpha=t}} = \frac{k_{eq} \pi e \phi_0}{2 K\left(\zeta\right) \sqrt{m_e \phi_0 e}}, \refstepcounter{equation}	  \label{eq:eqfreqa}	\subeqn \\
&\Omega_{\alpha=p}=\frac{k_{eq} \pi e \phi_0 \sqrt{\zeta}}{K\left(\zeta^-1\right) \sqrt{m_e \phi_0 e}},	   \label{eq:eqfreqb}   \subeqn
\end{align}
The behavior of these frequencies is similar to the bounce or transit frequency of the guiding center motion in tokamaks \cite{white}. It should be noted that \fref{fig1} is constructed by plotting $ \Omega_{\alpha} / \Omega _{\alpha} \left (  \zeta=10^{-3}   \right ) $.

\subsection{The linear growth rate}
\label{lgr}
In this subsection, we investigate the linear interaction between the plasma mode and the fast particles that are trapped and co/counter-passing in the electrostatic potential. For a traveling wave solution, the general form of the physical quantities can be represented as $ U = \sum_{n=1}^{\infty} \frac{e\phi_{n}}{2} \exp \left [ in \left ( k_{p} z-\omega t \right ) \right ] + c.c   = \sum_{n=1}^{\infty} \frac{e\phi_{n}}{2}   \sum_{p=-\infty}^{\infty} V_{\alpha,n,p} \left ( J_{\alpha} \right ) \exp \left [ i \left ( p\theta-n\omega t \right ) \right ] + c.c $ , $ \tilde{f}_{\alpha}=\sum_{n=1}^{\infty} \sum_{p=1}^{\infty}\hat{f}_{\alpha,n,p} \left ( J_{\alpha} \right ) \exp \left [i \left  ( p \theta - n \omega t \right ) \right ] + c.c$, $V_{c}=   \sum_{n=1}^{\infty}   \hat{V}_{n}\exp\left [ in\left (k_{p}z-\omega t\right )\right ]$, where $\omega=\omega_{r}+i \gamma_{l}$ is the complex frequency, $k_{p}$ the wave-number of the plasma mode, $V_{\alpha,n,p} \left ( J_{\alpha} \right )$ the orbit averaged mode amplitude which specifies the coupling strength and plays the same role as the so-called matrix element in \cite{berk1995,breizman1997},given by
\begin{equation}
V_{\alpha,n,p}=\frac{1}{2\pi}\int_{-\pi}^{\pi}\exp \left (ink_{p} z \right ) \exp \left (-ip \theta \right ) d\theta.
\label{eq:V}
\end{equation} 
In the previous BOT models for long range chirping \cite{Boris2010,Nyqvist2012,Nyqvist2013}, $V_{\alpha,n,p}$ is unity for the dominant resonance and is zero otherwise. In contrast, the presented approach enables investigation of different types of resonances in wave-particle interaction through a 1D model. It is noteworthy that the value of $\frac{k_p}{k_{\text{eq}}}=m$, where $m$ is an integer, can be associated with the mode numbers in realistic geometries. 

The total Hamiltonian describing the fast particle motion can be written in the form, $H_{\alpha}=H_{0,\alpha}+U$. This Hamiltonian along with the linearization of equation \eqref{eq:systemb}, is used to derive the linearized Vlasov equation in the form given by
\begin{equation}
\pdv{\tilde{f}_\alpha}{t} + \pdv{\tilde{f}_\alpha}{\theta} \pdv{ H_{0,\alpha}}{ J_\alpha} = \pdv{F_{\text{eq},\alpha} \left ( J_\alpha \right )}{J_\alpha}  \pdv{U}{\theta}.
\label{eq:linearizedvlasov}
\end{equation}      
Neglecting the higher harmonics $\left (n\geq 2 \right)$ in the linear approximation,
\begin{equation}
\hat{f}_{\alpha,n=1,p} = \frac{pe \phi_{1} V_{\alpha,n,p} \left (J_{\alpha}\right) \pdv{F_{\text{eq}} \left (J_\alpha \right )}{J_\alpha}}{2 \left (p\Omega_{\alpha} - \omega \right )}.
\label{eq:fpert}
\end{equation} 
It can be infered from expression \eqref{eq:fpert} that the resonance condition is 
\begin{equation}
\omega_{r}=p\Omega_{\alpha}.
\label{eq:res} 
\end{equation} 
The sign of $\Omega_{\alpha}$ is affected by the definition of the angle and considering $\Omega_{\alpha} > 0$, the resonance condition will be satisfied only for $p > 0$.
The perturbed density of the cold electrons can be derived from the linear fluid equations, \eqref{eq:systemc} and \eqref{eq:systemd}. To first order in perturbations, we have
\begin{align}
V_{c}=\frac{k_{p}U}{\omega m}, \refstepcounter{equation}	\label{eq:lfa}	\subeqn \\
\delta n = \frac{k_{p}^{2} n_{0} U} {m_{e} \omega^{2}}.  \label{eq:lfb}   \subeqn
\end{align}
Now we substitute the relevant terms into (\ref{eq:systema}) to find the dispersion relation of the mode given by
\begin{eqnarray}
\frac{\epsilon_{0} k_{p} m_{e}}{e^2} &\left ( 1- \frac{\omega_{pe}^{2}}{\omega^{2}} \right ) = \nonumber \\ &\sum_{\alpha} \int \sum_{p} \frac{p\left (\pdv{ F_{\text{eq},\alpha}}{J_{\alpha}} \right ) } { p\Omega_{\alpha}-\omega} \abs{V_{\alpha,p}}^2 dJ_{\alpha},
\label{eq:dispersion}
\end{eqnarray}
where $\omega_{\text{pe}}=\sqrt{ \frac {n_0 e^2} {m_e \epsilon_0} }$ is the electron plasma frequency. Neglecting the small contribution of the principal value which modifies the real part of the frequency inconsiderably, allows us to set $ \omega_r = \omega_{\text{pe}} $. Assuming $\gamma_{l} \ll \omega_{\text{pe}}$ (the wave evolves slowly compared with $\omega_{p}^{-1}$), equation \eqref{eq:dispersion} can be solved for $\omega $. Consequently, the linear growth rate is found to be
\begin{eqnarray}
\gamma_{l}=\frac{\omega_{\text{pe}} \pi e^{2}}{2\epsilon_{0}k_{p}m_{e}} \sum_{\alpha}\sum_{p} \pdv{F_{\text{eq},\alpha}}{\zeta_{\alpha}} \abs{V_{\alpha,p}}^2  \abs{\dv{\Omega_{\alpha}}{\zeta_{\alpha}}}
_{\Omega_{\alpha} \left ( J_{\alpha} \right ) = \frac{\omega_{\text{pe}}}{p} }^{-1}
\label{eq:lineargrowthrate}
\end{eqnarray}
which involves summing the contribution from all the resonances denoted by p. Equation \eqref{eq:lineargrowthrate} is a function of the energy parameter $\left ( \zeta \right )$. This indicates the dependency of the linear growth rate on particle orbits (Cf. \fref{fig3}). It should be noted that the contribution from the counter-passing electrons in the equilibrium potential is much less than the co-passing ones. This can be shown by changing $z$ to $-z$ in equation \eqref{eq:V} and evaluating the corresponding values of coupling strength for counter-passing electrons.

\subsection{Nonlinear BGK modes}
\label{bgk}

In the absence of collisions, the presence of any amount of dissipation leads to the formation of an unstable plateau in the distribution function of the energetic electrons which supports sideband oscillations that finally evolve into chirping modes \cite{Lilley2014}. The time scale of the motion of developed holes and clumps is much longer than the time scale of particles motion when they are trapped in the BGK mode, i.e. $ \dv{\omega_{b}}{t} \ll \omega_{b}^{2}$. Considering this adiabatic regime for the motion of phase-space structures after saturation of the mode amplitude, the kinetic equation can be bounce-averaged to find the perturbed distribution function of the fast electrons.   

Adopting a Fourier expansion for the periodic structure, the electrostatic energy of the nonlinear BGK mode can be written in the form
\begin{equation}
U[z,t]=\sum_{n} A_{n}(t) \cos \left [n \left (k_{p} z-\phi\left (t\right ) \right ) \right ],
\label{eq:BGKenergy}
\end{equation}
where the Fourier coefficients $A_{n}(t)$ evolve on a slow time scale but the periodic behavior of the BGK mode represents rapid oscillations with a time scale on the order of the inverse initial plasma frequency. The motion of the fast electrons can be investigated using the following Hamiltonian
\begin{eqnarray}
H_{\alpha}&= H_{\alpha,0}\left (J_{\alpha}\right )+ \frac{1}{2} \sum_{n} \sum_{p} A_{n}\left (t\right ) \nonumber \\
& \times V_{\alpha,n,p} \exp \left [  i \left ( p\theta - n\phi\left (t\right ) \right )  \right ] + c.c,
\label{eq:totalhamiltonian}
\end{eqnarray}
written in terms of the action--angle variables of the unperturbed motion. A simple canonical transformation can be used to cancel the fast time scale included in $\phi \left (t \right )$. We consider $\tilde{\theta}_{l}=l\theta-\phi \left (t\right )$ and $ \tilde{J}_{\alpha} = \frac{J_{\alpha}}{l}$ and the generating function for this canonical transformation is $\Phi\left [  \theta , \tilde{J}_{\alpha} , t \right ] = l\theta \tilde{J}_{\alpha} - \phi \left (t\right ) \tilde{J}_{\alpha}$, where $l=\frac{p}{n} $ denotes the type of the resonance. Considering the first resonance as having the dominant contribution to the interaction, the model can be evaluated by setting $l=1$. In section \ref{sec:result}, it is discussed that the contribution from the first resonance is dominant in this model. However, other types of resonances can be treated likewise. The new Hamiltonian is
\begin{eqnarray}
K_{\alpha}(\tilde{\theta}, \tilde{J}, t) &= H_{0,\alpha} \left ( \tilde{J}_{\alpha} \right ) - \dv{\phi \left (t \right )}{t} \tilde{J}_{\alpha} + \nonumber \\
&\frac{1}{2} \sum_{n} A_{n}\left (t\right ) V_{\alpha,n,n} \exp \left (in\tilde{\theta} \right ) + c.c.
\label{eq:newhamiltoniana}
\end{eqnarray}
The small separatrix width assumption allows us to neglect the higher order terms in the Taylor expansion of the unperturbed Hamiltonian near the resonant orbit. In addition, we also approximate $V_{\alpha,n,n}\left (\tilde{J} \right )$ with the first term of its Taylor expansion about $J_{\alpha,res}$. Using $ \left.\pdv{H_{\alpha,0}}{J_{\alpha}} \right |_{J_{\alpha}=J_{\alpha,res}\left (t\right )} = \Omega_{\alpha} = \dv{\phi\left (t\right )}{t} = \omega \left (t\right ) $ , the new Hamiltonian becomes
\begin{eqnarray}
K_{\alpha} =& \frac{1}{2} \left. \pdv[2]{H_{0,\alpha}}{\tilde{J}_{\alpha}}  \right |_{\tilde{J}_{\alpha} = J_{res,\alpha} \left (t \right )}  \left (\tilde{J}_{\alpha}-J_{res,\alpha}   \left (t\right ) \right )^2 +  \nonumber \\
&\frac{1}{2} \sum_{n} A_{n}\left (t \right ) V_{\alpha,n,n} \exp \left (in\tilde{\theta} \right ) + c.c.
\label{eq:newhamiltonianb}
\end{eqnarray}
\begin{figure}[b]
   \includegraphics[scale=0.6]{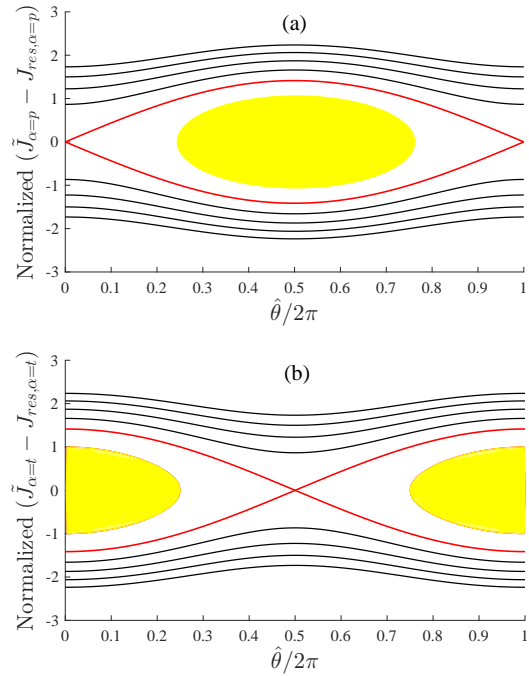}
  \caption{The panels (a) and (b) describe energy contours in phase space of the electrons which are passing and trapped in the equilibrium potential, respectively. The black lines specifiy the trajectories of the passing electrons and the shaded area is a sample of the adiabatic invariant of the trapped electrons in the nonlinear BGK mode. The red line is the separatrix.}
\label{fig2}
\end{figure}
Substituting $K_{\alpha}$ with the extremum value of the BGK mode electrostatic energy in equation \eqref{eq:newhamiltonianb}, gives the dynamics of the fast electrons on the separatrix supported by the nonlinear mode. This condition is used to identify the boundary of the trapped and passing electrons in the BGK mode, i.e. the separatrix. A simple manipulation of equation \eqref{eq:newhamiltonianb} gives
\begin{eqnarray}
\tilde{J}_{\alpha, \pm } & - J_{res,\alpha} \left (t \right ) = \nonumber \\ 
&\pm \left [ \left ( U_{\text{ext},\alpha} - \frac{1}{2} \sum_{n} A_{n} \left (t \right ) V_{\alpha,n,n} \left (J_{\alpha,res} \right )   \right. \right.  \nonumber \\
&\left. \left. \times \exp (in\tilde{\theta} ) + c.c \right ) \frac{2}{\Delta_{\alpha}} \right ] ^{ \frac{1}{2} },
\label{eq:separatrixwidth}
\end{eqnarray}
where $U_{\alpha,ext}$ is the extremum value of the BGK mode energy. The value of  $ \left. \pdv[2]{H_{0,\alpha}}{\tilde{J}_{\alpha}} \right |_{\tilde{J}_{\alpha} = J_{\text{res},\alpha}  \left (t\right ) } = \left.  \pdv{\Omega_{\alpha}}{\tilde{J}_{\alpha}} \right |_{\tilde{J}_{\alpha}=J_{\text{res},\alpha}  \left (t\right ) }$ (denoted by $\Delta_{\alpha}$) can be negative or positive for the trapped or passing electrons in the equilibrium potential, respectively. Mathematially, this affects $U_{\text{ext},\alpha}$ in order to have a positive value under the square root in equation \eqref{eq:separatrixwidth} and from the physical point of view, it shows that the passing electrons in the equilibrium potential are trapped in the energy well of the BGK mode, while the trapped electrons in the equilibrium potential are trapped in the energy hill of the BGK mode. This implies
\begin{equation}
\label{eq:Uext}
   U_{\text{ext},\alpha}=
    \begin{cases}
      U_{\text{min}}, & \alpha=t \\
      U_{\text{max}}, & \alpha=p 
    \end{cases}.
  \end{equation}
Phase-space trajectories of constant energy for the motion of energetic particles in the BGK mode are plotted in \fref{fig2}. It is shown that the separatrix supported by the nonlinear mode corresponding to the electrons trapped in the electrostatic potential (\fref{fig2}(\textit{b})) has a phase shift of $\pi$ with respect to the separatrix related to the passing group (\fref{fig2}(\textit{a})).

As the separatrix moves adiabatically, the phase-space area enclosed by the trajectories of the deeply trapped particles in the nonlinear wave, i.e. the shaded areas in \fref{fig2}, is conserved. Without trapping or detrapping over this region, the aforementioned conservation ensures that the value of the distribution function is conserved. The separatrix moves the trapped electrons in the BGK mode while the passing electrons are affected through the direction of their motion \cite{Boris2010}. The adiabatic invariant of the motion of these electrons in the BGK mode reads (Cf. \ref{Appendix} for more details)   
\begin{eqnarray}
I_{\alpha}= & 2\int_{0}^{2\pi} \left [ \left ( K_{\alpha} - \frac{1}{2} \sum_{n} A_{n}(t) V_{\alpha,n,n}  \right. \right.  \nonumber \\ 
& \left. \left. \times \exp(ip\tilde{\theta}) + c.c \right ) \frac{2}{\Delta_{\alpha}} \right ]^{ \frac{1}{2} } d\tilde{\theta}.
\label{eq:adiabatic}
\end{eqnarray}
Substituting expression \eqref{eq:BGKenergy} into equation \eqref{eq:systema} gives
\begin{eqnarray}
- \sum_{n} & A_{n}(t)n^{2}k_p^{2} \cos\left [ n \left ( kz-\phi \left (t \right ) \right )  \right ] = -\frac{e^2}{\epsilon_{0}} \nonumber \\
& \times \left [\frac{1}{m_e} \sum_{\alpha} \int_{-\infty}^{\infty} \tilde{f}_{\alpha}(z,p_{z})dp_{z}+ \delta n \right ],
\label{eq:poissonsubstitution}
\end{eqnarray}
where $\delta n$ can be derived under the linear response assumption of the bulk electrons. Similar to subsection \ref{lgr}, we multiply equation \eqref{eq:poissonsubstitution} by $ \cos \left [ n \left (k_{p}z-\phi(t) \right ) \right ]$ and integrate over one wave-length. We also write all the physical quantities in the fast particle term in terms of the new action--angle variables $(\tilde{\theta},\tilde{J})$. After substituting the Fourier expansion of $\cos \left [ n \left ( kz(\tilde{\theta},\tilde{J})-\phi(t) \right ) \right ] $ and neglecting the highly oscillating terms one finds
\begin{eqnarray}
A_{n} & \left (t \right )= \frac{1}{2\pi k_{p} n_{0}} \left [ \frac{\omega^{2}}{n^{2}\hat{\omega}^{2}-1}\right ] \sum_{\alpha} \int_{0}^{2\pi} d\tilde{\theta} \int_{0}^{\infty} \left [ \tilde{f}_{\alpha} (\tilde{\theta},\tilde{J} ) \right. \nonumber \\
&\left. \times  V_{\alpha,n,n}\exp ( in\tilde{\theta}) + c.c \right ]  \abs{\mathbf{J}} d\tilde{J},
\label{eq:Fouriercoef1} 
\end{eqnarray}
where the Jacobian of the canonical transformation $(z,p_{z}) \leftrightarrow (\tilde{\theta} , \tilde{J} )$ is unity and $\hat{\omega}=\frac{\omega}{\omega_{\text{pe}}}$ is the normalized frequency with respect to the initial electron plasma frequency. In this model, the phase-space density of the fast electrons (the distribution function) is assumed to be the same inside the narrow shrinking separatrix supported by the BGK mode, the so-called top-hat model. The perturbed part of the fast electrons distribution function dominated by the trapped electrons inside the separatrix \cite{Boris2010} is calculated using the bounce averaging method described in \ref{Appendix},
\begin{eqnarray}
\label{eq:perturbedpart}
   \tilde{f}_{\alpha}= \nonumber \\
    \begin{cases}
      0, & $passing in BGK$\\
      F_{eq,\alpha}\left (   J_{res}\left (t=0\right )\right ) - F_{eq,\alpha}\left (J_{res}\left (t\right )\right ). & $trapped in BGK $
    \end{cases} \nonumber \\
  \end{eqnarray}
Using the above expression, equation \eqref{eq:Fouriercoef1} transforms into
\begin{eqnarray}
A&_{n}\left (t\right ) = \frac{\omega^{2}}{2\pi k n_{0}\left ( n^{2} \hat{\omega}^{2} -1\right )} \sum_{\alpha} \left [ F_{eq,\alpha}\left (t=0\right ) - F_{eq,\alpha}\left (t \right ) \right ] \nonumber \\
&\times \int_{0}^{2\pi} d\tilde{\theta} \left [V_{\alpha, n,n} \exp(in\tilde{\theta}) + c.c \right ] \Delta \tilde{J}_{\alpha, max} \left (\tilde{\theta} \right ) ,
\label{eq:Fouriercoef2} 
\end{eqnarray}
where $\Delta \tilde{J}_{\alpha, \text{max}} \left (\tilde{\theta} \right )$ is the maximum width of the separatrix. Using equation \eqref{eq:separatrixwidth}, we have
\begin{eqnarray}
A_{n} & \left (t\right ) =  \frac{\omega^{2}}{\pi k n_{0}\left ( n^{2} \hat{\omega}^{2}-1 \right )} \sum_{\alpha} \left [ F_{eq,\alpha}\left (t=0 \right ) - F_{eq,\alpha}\left (t \right ) \right ] \nonumber \\
&\times \int_{0}^{2\pi}  \left [  (  U_{\alpha, \text{ext}} -\frac{1}{2} \sum_{n} A_{n}\left (t \right ) V_{\alpha,n,n}  \exp \left ( in\tilde{\theta} \right )   \right. \nonumber \\ 
&\left.  + c.c \right ) \left. \frac{2}{\Delta_{\alpha}} \right ] ^ { \frac{1}{2}  }  \left [V_{\alpha, n,n} \exp \left ( in\tilde{\theta} \right ) + c.c \right ]   d\tilde{\theta}.
\label{eq:Fouriercoef3}
\end{eqnarray}
The above equation can be solved numerically to derive the mode structure. The numerical method used is presented in section \ref{sec:numeric}.

The trapped electrons in the BGK mode travel in phase-space together with the nonlinear mode. Depending on whether they are trapped or passing in the equilibrium potential their energy increases or decreases with decreasing frequency of the mode. Hence, formation of a hole in the distribution function of trapped particles in the equilibrium potential accompanies a clump in the distribution of passing ones and vice versa. The change in the trapped electrons perturbed potential energy is relatively small when the change in the phase velocity is greater than the width of the separatrix. More energy is released by the fast particles via the motion of the phase-space structures than in the process of their formation and the released energy during chirping should compensate the dissipated energy into the bulk. The total amount of power released corresponding to the change of the structure energy is given by
\begin{equation}
P_{r}= -\sum_{\alpha} N_{\alpha} \dv{E_{\alpha}}{t},
\label{eq:releasedpower}
\end{equation}
where $N_{\alpha}$ is the total number of each group of electrons in the hole/clump, $ \dv {E_{\alpha}}{t}=\Omega_{\alpha} \left ( \dv{\Omega_{\alpha}}{J_{\alpha}} \right )^{-1}\dv{\omega \left (t\right )}{t}$ is the rate of change of the energy of each particle and the resonance condition allows setting $\Omega_{\alpha}=\omega \left(t\right)$. Regarding to the definition of the adiabatic invariant of the trapped particles, $N_{\alpha}$ can be calculated as
\begin{eqnarray}
N&_{\alpha} = \frac{2}{m_e} \left [  F_{\text{eq},\alpha}\left (t=0\right ) - F_{\text{eq},\alpha}\left (t \right ) \right ] \nonumber \\ 
&\times \int_{0}^{2\pi}  \left [ \left (     U_{\alpha, \text{ext}} -\frac{1}{2} \sum_{n} A_{n}\left (t\right ) V_{\alpha,n,n}  \right. \right. \nonumber \\   
 &\times \left. \left. \exp(in\tilde{\theta}) +c.c \right ) \frac{2}{\Delta_{\alpha}} \right ]^ {\frac{1}{2} } d\tilde{\theta}.
\label{eq:totalnumber} 
\end{eqnarray}
The work done by the collision force can be used to calculate the dissipated power $\left( P_{d}\right)$ into the bullk via collisions. Using the equation of motion \eqref{eq:systemc} and considering the collisional term, we have
\begin{equation}
P_{d} = \frac{2\pi \nu k_{p}}{\omega^{2}m_{e} } \langle U^{2} \rangle,
\label{eq:dissipatedpower} 
\end{equation}
where $\langle \rangle$ denotes averaging over one wavelength and $\langle U^{2} \rangle=\frac{1}{2} \sum_{n} A_{n}^{2} \left( t \right)$. The released power during the motion of the holes/clumps is equal to the power dissipated in the bulk through collisions. This power balance can be used to calculate the rate at which sweeping occurs, which results in
\begin{equation}
\dv{\omega \left (t\right )}{t} = - \left [  \frac{\nu n_{0} \pi k_{p}}{\omega^{3} m_{e}} \sum_{n} A_{n}^2 \left(t\right)    \right] \frac{1}{\sum_{\alpha} N_{\alpha} \left ( \dv{\Omega_{\alpha}}{J_{\alpha}} \right )^{-1}}
\label{eq:srate} 
\end{equation}

\section{Numerical Scheme}
\label{sec:numeric}
In this section, we first derive the equation of the mode structure at early state of chirping, say $ t_{0} $, considering only the contribution from the trapped electrons in the equilibrium potential. The infinitesimal imaginary part of the orbit averaged mode amplitude allows us to set $V_{\alpha,n,n} = \Re {(V_{\alpha,n,n})} $. Equation \eqref{eq:Fouriercoef3} states that at initial phase of sweeping only the first Fourier coefficient is non-zero (a sinusoidal mode structure) and is presented by
\begin{equation}
A_{1,0} = - \left [ \frac{8\omega_{\text{pe}}^{2} \pdv{F_{\text{eq},t}}{\zeta_{t}} \left. \pdv{\zeta_{t}}{\hat{\omega}} \right |_{\tilde{\omega}=1} }   {3\pi k_{p} n_{0} \sqrt{\abs{\Delta_{t,0}}}} \right ] V_{t,1,1,0} \sqrt{A_{1,0}V_{t,1,1,0}}.
\label{eq:initialcoef1} 
\end{equation}
Here, we have used the subscript 0 to denote evaluation at $t=t_{0}$. The term $A_{1,0}$ can be expressed in terms of the linear growth rate to have
\begin{equation}
A_{1,0} = \frac{16^2 \gamma_{l}^2 } {9\abs{\Delta_{t,0}} V_{1,0} \pi^4 } .
\label{eq:initialcoef2}
\end{equation}
We also let $\hat{A}_{n} \left (t\right )=A_{n}\left (t\right )/A_{1,0}$, $\hat{V}_{\alpha,n,n} \left (t\right )=V_{\alpha,n,n}\left (t\right ) / V_{t,1,1,0}$, $\hat{\Gamma}_{\alpha} = \Delta_{\alpha} / \abs{\Delta_{t,0}}$, $ \hat{U}_{\alpha,\text{ext}} = U_{\alpha,\text{ext}} / A_{1,0}V_{t,1,1,0}$ and $F_{eq,\alpha} \left (t \right ) = c_{\alpha} \zeta_{\alpha} \left (t \right ) $.
Normalizing equation \eqref{eq:Fouriercoef3} with respect to $A_{1,0}$ results in
\begin{eqnarray}
\hat{A}_{n} & \left (t \right ) = \left [ \frac{-3\hat{\omega}^2} {8c_{t} \left. \pdv{\zeta_{t}}{\hat{\omega}} \right |_{\hat{\omega}=1}   \left (  n^2 \hat{\omega}^2  - 1  \right ) } \right ]  \sum_{\alpha} c_{\alpha} \left [\zeta_{\alpha,0} -\zeta_{\alpha} \right ] \nonumber \\
&\times  \int_{0}^{2\pi}  \left [ \left ( \hat{U}_{\alpha,\text{ext}} - \frac{1}{2} \sum_{n} \hat{A}_{n}\left (t\right )\hat{V}_{\alpha,n,n} \exp \left ( in\hat{\theta} \right )  \right. \right.    \nonumber \\
&\left. \left.  + c.c \right )   \frac{2}{\hat{\Gamma}_{\alpha}} \right ]^{ \frac{1}{2} }      \left [ \hat{V}_{\alpha,n,n} \left (t\right ) \exp \left ( in\hat{\theta} \right ) + c.c \right ]   d\hat{\theta},
\label{eq:finalcoef}
\end{eqnarray}
which can be solved iteratively to derive the Fourier coefficients. In order to avoid the singularity in the numerical approach, a special treatment is applied to the first coefficient when the values of $\omega$ are close to $\omega_{\text{pe}}$. In this case, $\zeta_{\alpha}\left (t \right )$ can be linear-approximated around the initial plasma frequency to cancel the effect of the pole in the denominator of equation \eqref{eq:finalcoef}. \\
Likewise, equation \eqref{eq:srate} can be investigated for the early phase of the structures motion in phase-space considering only the effect of trapped particles in the electrostatic potential. Substituting expression \eqref{eq:totalnumber} into equation \eqref{eq:srate} and using equations \eqref{eq:lineargrowthrate} and \eqref{eq:initialcoef2}, one finds
\begin{equation}
\dv{}{t} \frac{\left (\omega-\omega_{pe}\right)^{2}}{\omega_{pe}^{2}} = \frac{\nu}{3} \left ( \frac {16\gamma_{l}}{3 \pi^{2} \omega_{pe}} \right )^{2}.
\label{eq:initialsrate}
\end{equation}
We define the dimensionless time $\tau = \frac{\nu}{3}\left ( \frac{16\gamma_{l}}{3\pi^{2} \omega_{\text{pe}}} \right )^{2}t   $ and multiply equation \eqref{eq:srate} by $ \frac{3}{\nu \left( \frac {16 \gamma_{l} }{3 \pi^{2} \omega_{\text{pe}}} \right)^{2}} $ to have
\begin{eqnarray}
\dv{\hat{\omega}}{\tau} =& - \left [ \frac{4}{\hat{\omega}^{3}} \right ] \frac{c_{t} \abs{\left ( \dv{\hat{\omega}}{\zeta}  \right )_{t,0}^{-1}}_{\zeta=\zeta_{\text{resonance}}} \sum_{n} \hat{A}_{n}^{2} }   { \sum_{\alpha} \text{sgn}_{\alpha} c_{\alpha} \left [ \zeta_{\alpha,0} - \zeta_{\alpha} \right ]  }  \nonumber \\
&\times \left \{ \int_{0}^{2\pi} \left [ \left ( \hat{U}_{\alpha,\text{ext}} - \sum_{n} \frac {\hat{A}_{n} }{2} \hat{V}_{\alpha,n,n} \right. \right. \right.  \nonumber \\ 
&\times   \left.  \left.  \left. \exp \left (in \hat{\theta} \right ) + c.c \right )    \frac{2}{\hat{\Gamma}_{\alpha}^{3}} \right ]^{ \frac{1}{2} }   d\hat{\theta}  \right \}^{-1}  
\label{eq:finalsweepingrate}
\end{eqnarray}
where $\text{sgn}_{\alpha}$ is -1 and 1 for $\alpha=t$ and $p$, respectively. The above equation can be solved by a fourth-order Runge-Kutta method along with the iterative method used for solving the Fourier coeffcients on the RHS.

In case that the energy of the electrons is high enough with respect to the electrostatic potential energy (deeply passing electrons with $ \zeta \gg 1 $ ), their motion will not be affected by the equilibrium potential and they move freely. In other words, $\theta=k_{\text{eq}}z$. Subsequently, only one resonance is non-zero and the orbit averaged mode amplitude is equal to unity (Cf. \fref{fig4}(\textit{b})) under this condition. In this high energy range, one can find that $k_{p}z =p \theta$ in the linear theory limit. Canonical equations of motion assure $\theta=\Omega_{\alpha=p}t$ so using equation \eqref{eq:res}, the resonance condition becomes $\omega=k_{p}v$, where $v$ is the particle velocity. Consequently, solving equations \eqref{eq:finalcoef} and \eqref{eq:finalsweepingrate} in the limit that $ \zeta \gg 1 $,  reproduces exactly the same results as in \cite{Boris2010}, which serves as the benchmark of the code and the numerical approach. 

\section{Results}
\label{sec:result}

\begin{figure}[h]
\centering
   \includegraphics[scale=0.6]{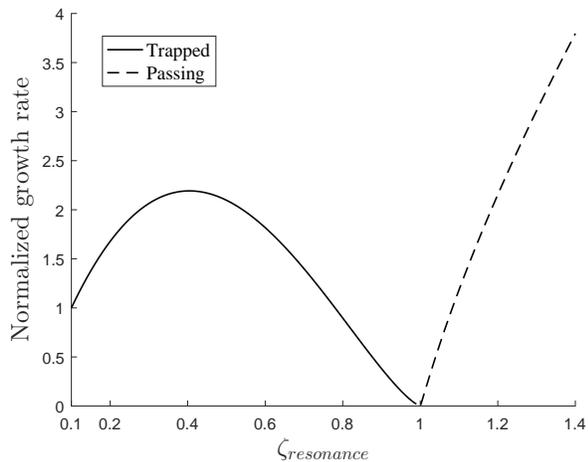}
  \caption{The linear growth rate behavior, corresponding to the first resonance, for different energy ranges of trapped and passing particles in the potential. All the values are normalized to the value at $\zeta=0.1$.}
  \label{fig3}
\end{figure}

For illustration, we have arbitarily restricted attention to cases where $ k_{\text{p}}=k_{\text{eq}} $ and the equilibrium potential has fixed amplitude. In the linear regime, the plasma mode will grow at different rates depending on the initial orbits of the electrons interacting with the mode. \Fref{fig3} demonstrates that the linear growth rate decreases to zero in the limit of having resonance with the particles close to the separatrix in the equilibrium potential.
\begin{figure}[t]
\centering
   \includegraphics[scale=0.66]{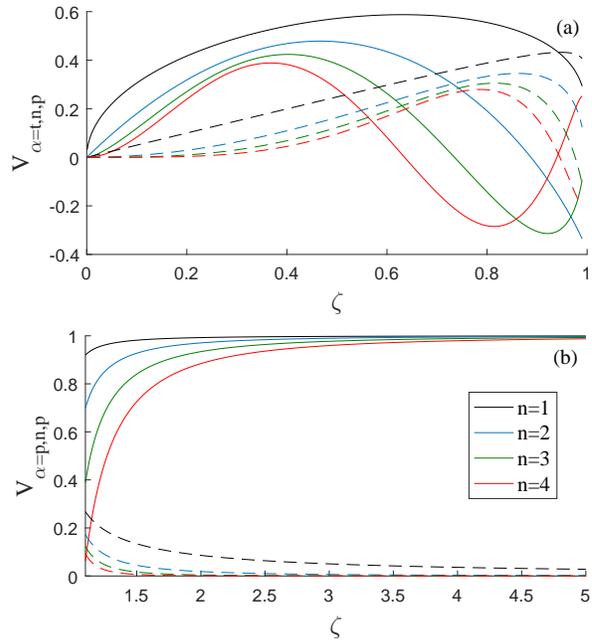}
  \caption{The orbit averaged mode amplitude versus energy parameter for (a) the trapped and (b) passing electrons in the equilibrium potential. The solid and dashed lines correspond to first and second resonances, respectively.}
  \label{fig4}
\end{figure}

As in subsection \ref{bgk}, the first resonance (l=1) is considered as the dominant resonance contributing to the interaction. The first four elements of the orbit averaged mode amplitude $ \hat{V}_{\alpha,n,p} $, indicating the coupling strength, corresponding to the first ($ \hat{V}_{\alpha,n,n} $) and the second ($ \hat{V}_{\alpha,n,2n} $) resonances are plotted in \fref{fig4} versus energy parameter. Investigation of \fref{fig4} shows that there are regions (adjacent to $ \zeta=1 $) where the values of the dominant element (n=1) belonging to the second resonance overtake the values of the dominant element of the first resonance. In itself, this may indicate that the corresponding second resonance is dominant. However, consideration of the growth rate for different resonances shows that the first resonant $\left (p =1 \right ) $ is dominant. This can be understood by inspection of equation \eqref{eq:lineargrowthrate}: the term $ \abs{\dv{\Omega_{\alpha}}{\zeta_{\alpha}}} $ increases with increasing the resonance, so $\gamma_l$ decreases with increasing resonance. In addition, evaluating the factors of equation \eqref{eq:finalcoef} for higher resonances $\left (l \geq 2 \right ) $ shows that its always the first resonance $\left (l=1 \right ) $ that has dominant contribution to the interaction in the hard nonlinear regime. Therefore, the submissive resonances are neglected.
The other important point about the coupling strength is that all of its elements go asymptotically to zero as the energy parameter of the electrons approaches unity. Here, we explain this phenomenon in more detail: Considering the canonical transformation used in subsection \ref{sec:dynamics}, the equations describing the position $(z)$ of the particle in terms of the action--angle variables, read
\begin{align}
z_{\alpha=t}  = \frac{2}{k_{\text{eq}}} \sin ^ { -1 } \left [ \sqrt{\zeta} \text{Sn} \left ( \frac{2 \theta K \left ( \zeta \right )}{\pi}, \zeta \right ) \right ], \refstepcounter{equation}	\label{eq:za}	\subeqn \\
z_{\alpha=p} = \frac{2}{k_{\text{eq}}} \sin ^ { -1 } \left [ \text{Sn} \left (\frac{\theta K \left ( \zeta ^ -1 \right )}{\pi}, \zeta ^-1 \right ) \right ],  \label{eq:zb}  \subeqn 
\end{align}
where Sn is the Jacobi elliptic function.
\begin{figure}[t]
\centering
   \includegraphics[scale=0.5]{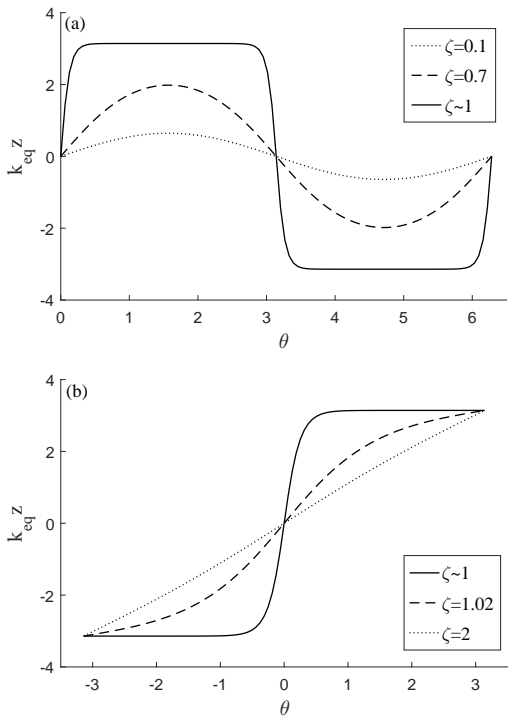}
  \caption{The position of energetic electrons (a) trapped and (b) passing in the equilibrium potential in terms of the angle variable}
  \label{fig5}
\end{figure}
\Fref{fig5} shows the position of the electrons in the electrostatic potential well at different times for different energy parameters. For the case of trapped (\fref{fig5}(\textit{a})) and passing (\fref{fig5}(\textit{b})) electrons, it can be shown that for $ \zeta \approx 1 $, the electrons spend most of their period lingering at the two peaks on the top of the well. In fact, due to energy conservation for a specific orbit in the unperturbed motion, the kinetic energy decreases to zero at these two points where the potential energy of the particle becomes maximum. Therefore, the value of the coupling strength reaches zero for $\zeta \approx 1 $, similar to the case where the electrons are deeply trapped $ \left ( \zeta \approx 0 \right ) $.

Prior to solving the equations for the mode structure and the sweeping rate in the hard nonlinear regime, it is necessary to investigate the behavior of the adiabatic invariant (phase-space area) of the trapped electrons in the BGK mode that are trapped or passing in the electrostatic potential. 
\begin{figure}[t]
\centering
   \includegraphics[scale=0.54]{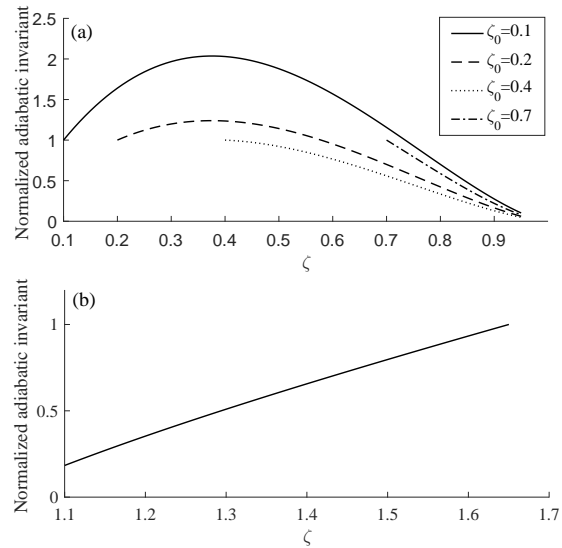}
  \caption{The values of the adiabatic inavariant of (a) trapped and (b) passing particles in the equilibrium potential for $\frac{k_{p}}{k_{\text{eq}}}=1$ at the separatrix determined by the BGK mode. The values are normalized with respect to the values at initial phase of chirping.}
  \label{fig6}
\end{figure}
\Fref{fig6} shows the values of the adiabatic invariant (equation \eqref{eq:adiabatic}) at the separatrix determined by the BGK mode during frequency sweeping. For the case of downward frequency sweeping, the energy of the passing electrons in the equilibrium potential decreases, so does the corresponding value of the adiabatic invariant (Cf. \fref{fig6}(\textit{b})). However, for trapped electrons, energy increases for downward frequency sweeping. Depending on the initial orbit, the adiabatic invarinat can either initially increase ($\zeta < 0.4$ of \fref{fig6}(\textit{a})) or decrease ($\zeta \geq 0.4$ of \fref{fig6}(\textit{a})). Due to the assumption of a flat-top distribution function over the separatrix region, the model remains valid as long as the separatrix supported by the BGK mode shrinks and an expanding separatrix (an increasing adiabatic invariant) should be avoided. Therefore, the electrons in the following results have initial energies parameters $ \zeta \geq 0.4 $. In this range, the coherent phase-space structure is a hole whose separatrix area (and the correspoding amplitude of the mode) is shrinking for a downsweeping frequency. For the case that new electrons are trapped into an expanding separatrix, it is required that the value of the distribution function of newly trapped particles is set to the value of the ambient distribution. The latter case is not the subject of this paper.

\subsection{The mode structure}
\label{subsec:modestructure}

\begin{figure*}
  \centering
\includegraphics[scale=0.525]{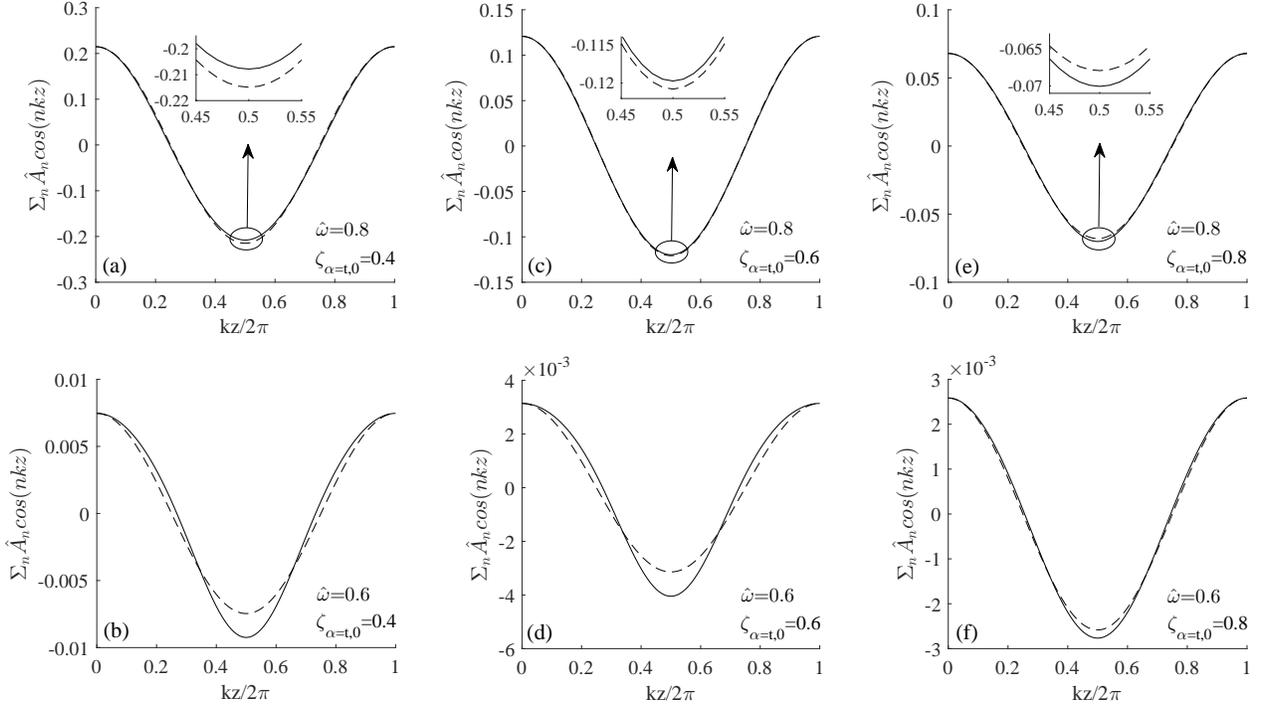}
\caption{The normalized BGK mode structure affected by electrons having different initial energies. The dashed line, included here for comparison, represents the sinusoidal structure of the mode at early stage of frequency sweeping}
\label{fig7} 
\end{figure*}
Considering similar slopes  for the initial distribution of both the trapped and passing electrons in the electrostatic potential (simultaneously in resonance with the plasma mode), the structure of the BGK mode has been solved for different initial electron energy parameters, namely $ \zeta_{\alpha=t}(t=0)$ = 0.4, 0.6 and 0.8. \Fref{fig7} illustrates the mode structure for these initial energies in cases where $ \hat{\omega} $= 0.8 and 0.6. The results reveal that for a nonzero change in $ \hat{\omega} $, the nonlinear behavior of the BGK mode is determined by the initial electron orbits. For constant $\hat{\omega}$, e.g. figures \ref{fig7}(\textit{a}), (\textit{c}) and (\textit{e}), the maximum amplitude of the mode structure (maximum value of $ \sum_n \hat{A}_n cos (n k_p z) $ ) changes with changing $\zeta_{\alpha=t,0} $, and the change in the mode amplitude decreases with increasing $\zeta_{\alpha=t,0} $. The shape of the nonlinear structure is not only affected by the amount of change in the frequency $\left (  \hat{\omega} \right )$ but also by the initial energy parameter $ \left (  \zeta_{\alpha=t,0} \right )  $. In order to explain the observed behavior, we first calculate the contribution of the trapped and passing particles to the mode structure seperately while they are simultanously in resonance with the mode. Afterwards, the behavior of both the equilibrium frequency and the physical quantities appearing in equation \eqref{eq:finalcoef} is investigated.

\begin{figure}[b]
	\centering
	\includegraphics[scale=0.6]{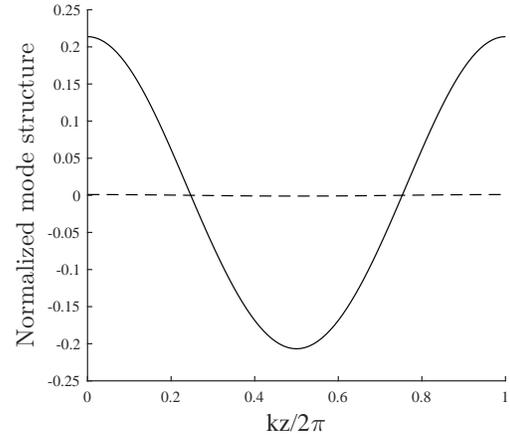}
	\caption{The contribution of trapped (solid line) and passing (dashed curve) electrons to the mode structure, where $\hat{\omega}=0.8$ and $\zeta_{\alpha=t,0}=0.4$. plotting on the same chart results in the small contribution of passing electrons to appear as a horizontal line}
  \label{fig8} 
\end{figure}
The Fourier coefficients are calculated by adding the two terms on the RHS of equation \eqref{eq:finalcoef}, corresponding to $ \alpha=\text{t}$ and $\text{p}$. The seperate contributions of these two groups of particles to the mode structure are shown in \fref{fig8} for similar values of distribution function and in case of simultanous resonance between the plasma mode and these two types of energetic particles orbit. It is clear that the contribution of the passing electrons to the nonlinear behavior of the mode is relatively much smaller than the trapped ones. The reason being that the resonance occurs in a region where the equilibrium frequency of passing particles has much steeper gradient in energy (Cf. \fref{fig1}). Therefore, for the purpose of investigating the parameters of equation \eqref{eq:finalcoef}, we only consider the dominant contribution from the trapped electrons in the fixed potential.

At a constant value of the normalized frequency $\hat{\omega}$, a simple evaluation of equation \eqref{eq:finalcoef} gives 
\begin{equation}
\hat{A}_{n}\left (t \right ) \propto \frac{ ( \dv{\hat{\omega}}{\zeta} )_{\alpha=t,0}^2 [ \zeta_{\alpha=t} \left (t=0\right )-\zeta_{\alpha=t} \left (t \right )]^2 \hat{V}_{\alpha=t,n,n}^3}{\hat{\Gamma}_{\alpha=t}}.
\label{eq:coefprop}
\end{equation}
Starting from different initial energies, the trapped electrons in the equilibrium potential should be moved on different energy increments by the nonlinear mode in order to have the same amount of change in the frequency. 
\begin{figure}[t]
\centering
   \includegraphics[scale=0.76]{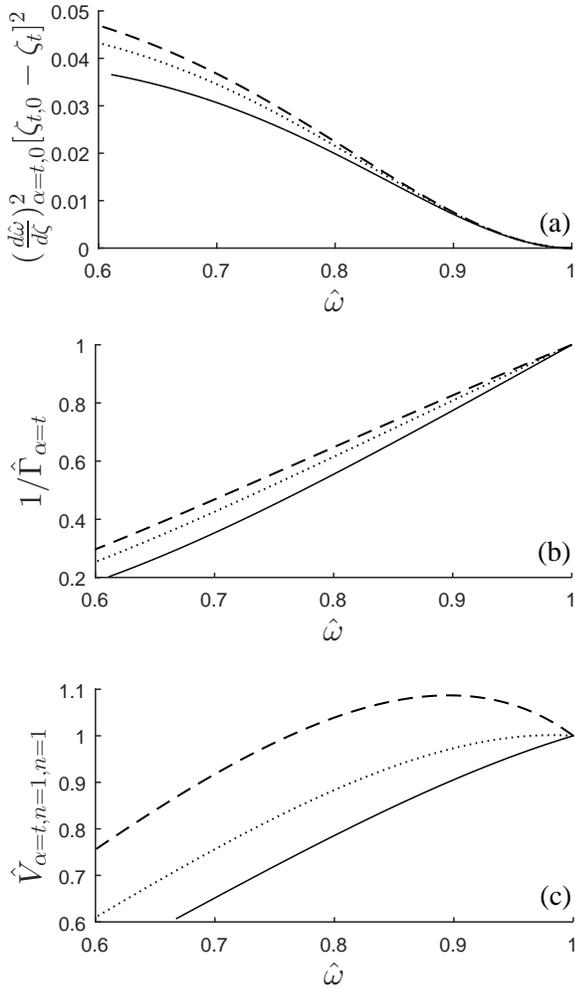}
  \caption{The factors in \eqref{eq:coefprop} versus the normalized frequency $\left ( \hat{\omega} \right ) $. The dashed, dotted and solid lines correspond to the initial energy parameter $\left ( \zeta_{\alpha=t,0} \right ) $  values of 0.4, 0.6 and 0.8, respectively.}
\label{fig9}
\end{figure}
This results from the nonlinear dependency of the equilibrium frequency on the energy parameter (Cf. \fref{fig1}). As an example for $\hat{\omega}=0.8$, the fast electrons having the initial energy parameters of $\zeta_{\alpha=t ,0}$=0.4,0.6 and 0.8 should be moved in phase-space to the points where $\zeta(t) $=0.783, 0.863 and 0.94, respectively and the energy increments become shorter for higher values of initial energy parameter. For a linear equilibrium distribution, the difference in the energy increments will explicitly appear in the nominator of equation \eqref{eq:finalcoef} through the perturbed density term, i.e. $ \left [\zeta \left ( t=0 \right ) - \zeta \left ( t \right ) \right ] $ . In general, the nonlinear dependency of the equilibiurm frequency on the energy parameter (Cf. \fref{fig1}) will affect the values of all the physical parameters apearing in equation \eqref{eq:finalcoef} for a fixed amount of frequency shift. \Fref{fig9} shows the dependency of the factors $ \left ( \dv{\hat{\omega}}{\zeta} \right )_{\alpha=t,0}^2 [ \zeta(t=0)-\zeta(t)]^2 $, $ \hat{V}^3 $ and $ \hat{\Gamma}^{-1} $ in \eqref{eq:coefprop} for different $\zeta_{\alpha=t,0}$ and as a function of $\hat{\omega} $. The dependency of $ \hat{A}_{n} $ with $\zeta_{\alpha=t,0} $ can be understood by inspection of these factors. At each $\hat{\omega}$ the factors decreases with increasing $\zeta_{\alpha=t,0}$ and so $ \hat{A}_{n} $ decreases.

\subsection{The sweeping rate}
\label{subsec:sweep}

In this subsection, we investigate the rate at which the frequency of the nonlinear mode evolves in time. Prior to solving the equation \eqref{eq:finalsweepingrate}, we evaluate the dependency of the sweeping rate $ \left ( \dv{\hat{\omega}}{\tau} \right ) $ on the initial energy parameter of the electrons (initial orbits) using the behavior of the factors illustrated in \fref{fig9}. Looking at the expression \eqref{eq:finalsweepingrate} for the sweeping rate at a constant value of $\hat{\omega}$, it can be inferred that
\begin{equation}
\dv{\hat{\omega}}{\tau} \propto \frac{\hat{A}_{n}(t)^{\frac{3}{2}}  \hat{\Gamma}_{\alpha=t} ^{\frac{3}{2}}} {\abs{\left(\dv{\hat{\omega}}{\tau}\right)_{\alpha=t,0}} [\zeta_{\alpha=t}(t=0)-\zeta_{\alpha=t}(t)] \hat{V}_{\alpha=t,n,n}^{\frac{1}{2}}}.
\label{eq:srateprop1}
\end{equation}
Using expression \eqref{eq:coefprop} one finds
\begin{equation}
\dv{\hat{\omega}}{\tau} \propto  \left ( \dv{\hat{\omega}}{\zeta} \right ) _{\alpha=t,0}^{2} [\zeta_{\alpha=t}(t=0)-\zeta_{\alpha=t}(t)]^2 \hat{V}_{\alpha=t,n,n}^4.
\label{eq:srateprop2}
\end{equation}

\begin{figure}[t]
\centering
   \includegraphics[scale=0.68]{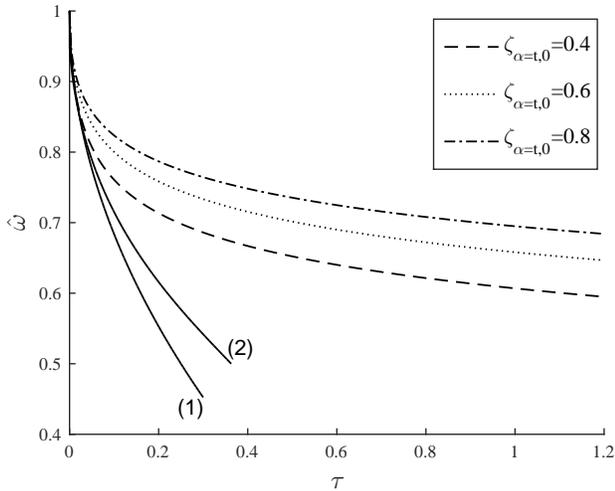}
  \caption{The evolution of normalized frequency versus normalized time. The solid lines labeled (1) and (2) correspond to the square root dependency, plotted for comparison, and the result reported in \cite{Boris2010}, respectively.}
\label{fig10}
\end{figure}

Similar to subsection \ref{subsec:modestructure}, one can consider figures \ref{fig9}(\textit{a}) and (\textit{c}) at a constant $ \hat{\omega}$ to investigate the value of the RHS of equation \eqref{eq:srateprop2} for different electron orbits. It is clear that the RHS value becomes lower when the resonance occurs with the electrons (trapped in the fixed equilibrium potential) having higher initial energy parameter $\left ( \zeta_{\alpha=t,0} \right )$. Therefore, we expect the mode frequency to chirp slower when the initial energy parameter of the electrons is higher. This can be verified by solving equation \eqref{eq:finalsweepingrate} using the numerical method stated in section \ref{sec:numeric} for different initial orbits. \Fref{fig10} illustrates the time evolution of $\hat{\omega}$ for different values of $\zeta_{\alpha=t,0}$. The results reproduce the square root dependency for initial stages of chirping as in \cite{Berk1997,Boris2010}. 
However, it is shown that in this model, the holes and clumps can move with much lower rates compared with the sweeping rates observed in \cite{Boris2010}. On the other hand, as predicted above, for higher initial energy parameter of the trapped electrons in the equilibrium potential, the frequency tends to decrease more slowly.

\section{Concluding remarks}
\label{sec:conclusion}

The more realistic 1D model shows that apart from the amount of deviation from the initial eigenfrequency during frequency sweeping, the initial orbit (initial energy parameter) of the particles in an equilibrium potential, determines both the linear and the hard nonlinear evolution behavior of a plasma mode. The model also resolves the simultaneous contributions from the two groups of particles having different orbit types as well as the contribution from higher resonances. We find however that the first resonance is dominant. We also identify different behavior of the adiabatic invariant in different energy regions. The model shows that for a constant trend in frequency sweeping, either upward or downward, the adiabatic invariant can have both positive and negative gradients in the energy parameter depending on the energy region considered. This behavior depends on factors such as the resonance number, the proportion of the plasma mode wave-number to the one for equilibrium potential ($\frac{k_{p}}{k_{\text{eq}}}$) and whether the particles were initially trapped or passing in the equilibrium potential. This indicates that for realistic geometries where particles interacting with the mode can follow different equilibrium orbits, an extended approach is required to calculate the perturbed density inside the holes and clumps. This extension can highly benefit from the method presented in \cite{Nyqvist2013}. However, it should be taken into account that the adiabatic invariant (phase-space area) at the separatrix can have both the shrinking and expanding behavior depending on the initial orbit of the energetic particles.

The presented model in this manuscript provides a more effective understanding of hard nonlinear wave-particle-plasma interactions in realistic geometries provided that the mode is subject to weak continuum damping (a global mode) i.e. its structure in the linear regime is not mainly determined by the energetic particles. Two different orbit topologies of energetic particles created by adding an electrostatic potential to the 1D bump-on-tail instability problem, bring it into anology with tokamaks where there exist trapped/passing topologies which can both resonate with modes with different coupling strength factors. In a high aspect ratio tokamak, the total magnetic field follows  
\begin{equation}
B \propto \frac{1}{R_{0} + r \cos\theta} \propto \frac{1}{R_{0}} \left(  1-\epsilon \cos\theta \right  ),
\end{equation}
where $B$ is the magnetic field, $\epsilon$ is the inverse aspect ratio, $\theta$ is the poloidal angle and $R_0$ and $r$ are the major and minor radius, respectively.

Using the orbit-averaged Littlejohn's Hamiltonian \cite{littlejohn}, we have 
\begin{equation}
H_{0} - \mu B_{0} = \frac{1}{2}m_{i}v_{\parallel}^{2} - \mu B_{0} \epsilon \cos \left ( \theta \right ),
\end{equation}
where $ H_{0}$ is the equilibrium Hamiltonian, $ \mu $ is the magnetic moment, $ m_{i}$ is the mass of the ion and $ v_{\parallel}$ is the velocity in the direction of the magnetic field.
Taking into account the symmetry of the magnetic field in toroidal direction in realistic geometries and assuming that the deviation of the fast particles from the flux surface is infinitesimal, the above Hamiltonian is comparable to the equilibrium Hamiltonian presented in equation \eqref{eq:equilibriumhamiltonian}. Further restrictions on the perturbation such as symmetry in toroidal direction, being localized on one flux surface and the assumption that the perturbation on different flux surfaces are unlinked, might let the presented model to describe some features of electrostatic axisymmetric modes ($n=0$ , where n is the poloidal mode number), namely global geodesic acoustic modes (GGAMs) in more realistic geometries \cite{Berk2006}. Nevertheless, an exact description of excited Global-Alfven-Eigenmodes (GAEs) with an evolving mode structure during long range frequency deviations requires the extension of the presented model, which is a part of our ongoing research.  

\section*{Acknowledgments}
The authors wish to thank Prof. Boris Breizman and Dr. Michael Fitzgerald for stimulating discussions that helped inspire this paper. This work was funded by the Australian Research Council through Grant No. DP140100790.

\section*{References} 

\bibliography{article}
\bibliographystyle{iopart-num}

\clearpage

\appendix

\section{Adiabatic invariant and bounce averaging method}
\label{Appendix}
The adiabatic invariant for a Hamiltonian $K(\hat{\theta},\hat{J}, \lambda \equiv \beta t) $ with slow time dependency ($\beta \ll $ typical orbit frequencies) is
\begin{equation}
I^{\infty} =  I \left (q,p,\lambda \right ) + \beta I_{1} \left ( q,p,\lambda \right ) + \beta^{2} I_{2} \left ( q,p,\lambda \right ) + ...,
\end{equation}
which the lowest term is commonly taken to be the action, $ I \left ( E , \lambda \right ) = \oint \hat{J} \left ( \hat{\theta},E,\lambda  \right ) d \hat{\theta} $ with $  K \left (\hat{\theta} , \hat{J} , \lambda  \right ) = E$. We transform to action-angle variables using the generating function $ \Phi_{2} \left ( \hat{\theta}, I , \lambda \right ) = \int_{\hat{\theta}_{0} \left ( I,\lambda \right )}^{\hat{\theta}}   d\hat{\theta}'  \hat{J} \left (  \hat{\theta}' ,  K \left ( I , \lambda \right ) , \lambda \right ) .$ So the Hamiltonian transforms into $ K_{new} \left ( \phi, I, \lambda \right ) = K \left ( I , \lambda \right ) + \beta \pdv{\Phi_{2}}{\lambda}.$ Now we consider the trapped electron Vlasov equation
\begin{equation}
\pdv{f}{t} + \pdv{f}{\phi}\pdv{K_{new}}{I} - \pdv{f}{I} \pdv{K_{new}}{\phi} = 0.
\label{A2}
\end{equation}
Using the equations of motion we have
\begin{align}
&\dot{\phi} = \pdv{K_{new}}{I}= \omega_{Bounce}+\pdv{}{I}\pdv{\Phi_{2}}{t}, \label{A3a} \refstepcounter{equation} \subeqn \\
&\dot{I} = \pdv{K_{new}}{\phi}  = \pdv{}{\phi} \pdv{\Phi_{2}}{t}   ,\label{A3b}  \subeqn 
\end{align}
Substituting the above expressions in equation \eqref{A2} gives
\begin{equation}
\pdv{f}{t} + \pdv{f}{\phi}\omega_{Bounce}+ \pdv{f}{\phi} \pdv{}{I} \pdv{\Phi_{2}}{t} - \pdv{f}{I}\pdv{}{\phi}\pdv{\Phi_{2}}{t} = 0.
\label{A4}
\end{equation}
Following the same approach in \cite{Nyqvist2012}, $f$ can be expanded in terms of the small parameter $\beta = \frac{\tau_{B}}{\tau_{s}}$ to have
\begin{equation}
f = f_{0} + \beta f_{1} + \beta^{2} f_{2} + ...,
\label{A5}
\end{equation}
where $f_{0}$ is the bounce average of $f$ over $\phi$. Using expression \eqref{A5}, we substitute for $f$ in equation \eqref{A4}. To lowest order $ \left ( \mathcal{O} \left (1\right ) \right ) $ in $\beta$, one finds
\begin{equation}
\pdv{f_{0}}{\phi}=0.
\label{A6}
\end{equation}
To next order $ \left ( \mathcal{O} \left ( \beta \right ) \right )$,
\begin{eqnarray}
\pdv{f_{0}}{t} & + \beta \pdv{f_{1}}{t} + \pdv{f_{0}}{\phi} \omega_{Bounce} + \beta \pdv{f_{1}}{\phi} \omega_{Bounce}     \nonumber \\ 
& +\pdv{f_{0}}{\phi} \pdv{}{I}\pdv{\Phi_{2}}{t} + \beta \pdv{f_{1}}{\phi} \pdv{}{I} \pdv{\Phi_{2}}{t} - \pdv{f_{0}}{I}\pdv{}{\phi} \pdv{\Phi_{2}}{t} \nonumber \\
& \ \ \ \ \ \  - \beta \pdv{f_{1}}{I} \pdv{}{\phi} \pdv{\Phi_{2}}{t} = 0.
\label{A7}
\end{eqnarray}
The second, sixth and eighth terms are on the order of $ \beta^{2} $ $ \left ( \mathcal{O} \left (\beta^{2} \right ) \right )$ and can be neglected at this stage. Equation \eqref{A6} shows that $f_{0} $ is independent of $\phi$, which allows us to set the fifth term to zero. Therefore, we reach

\begin{equation}
\pdv{f_{0}}{t} + \beta \pdv{f_{1}}{\phi} \omega_{Bounce} - \pdv{f_{0}}{I}\pdv{}{\phi} \pdv{\Phi_{2}}{t}  = 0.
\label{A8}
\end{equation}
After averaging \eqref{A8} over $\phi$, the second and third terms vanish and we find
\begin{equation}
\pdv{f_{0}}{t} = 0.
\label{A9}
\end{equation}
We define $f_{0} = \delta f + \left <  F_{\text{eq}} \left ( J_{\text{res}} \left ( t \right ) \right ) \right> $, where $<>$ denotes averaging over $\phi$ and $  f_{0} \left ( t=0 \right ) = F_{\text{eq}} \left ( J_{\text{res}} \left ( t=0 \right ) \right )$. The uniformity assumption of the distribution function over the separatrix region assures $\left <  F_{\text{eq}} \left ( J_{\text{res}} \left ( t \right ) \right ) \right>  =  F_{\text{eq}} \left ( J_{\text{res}} \left ( t \right ) \right ) $. Hence, $ f_{0} \left ( t \right ) = \delta f + F_{\text{eq}} \left ( J_{\text{res}} \left ( t \right ) \right )$. According to \eqref{A9}, $f_{0}$ should remain constant during frequency sweeping which gives
\begin{equation}
\delta f = F_{\text{eq}} \left ( J_{\text{res}} \left ( t=0 \right ) \right ) - F_{\text{eq}} \left ( J_{\text{res}} \left ( t \right ) \right ).
\label{A10}
\end{equation}
\end{document}